\documentclass{article}[12]
\usepackage[top=1in, bottom=1in, left=1.25in, right=1.25in]{geometry}
\usepackage{amsmath}
\usepackage[english]{babel}
\usepackage[utf8]{inputenc}
\usepackage{graphicx}
\usepackage{subcaption}
\usepackage[usenames, dvipsnames]{xcolor}
\usepackage{float}
\usepackage{hyperref}
\usepackage{amsmath}
\usepackage{amsthm}
\usepackage{ulem}
\usepackage{natbib}
\usepackage{lineno}
\usepackage{setspace}
\usepackage{amsfonts} 
\usepackage{authblk}
\usepackage{url}
\usepackage{amssymb} 
\usepackage{bm}
\usepackage{booktabs}

\newcommand{\Po}{\mathrm{Po}}

\newcommand{\igamma}{\mathrm{Inv.Gamma}}

\newcommand{\bsZ}{\boldsymbol{Z}}
\newcommand{\bsN}{\boldsymbol{N}}
\newcommand{\bsTheta}{\boldsymbol{\theta}}
\newcommand{\W}{\mathrm{W}}

\newcommand{\E}{\mathbb{E}}
\newcommand{\Var}{\mathrm{Var}}
\newcommand{\Cov}{\mathrm{Cov}}
\newcommand{\bstilZ}{\boldsymbol{\widetilde{Z}}}

\renewcommand{\d}{\mathrm{d}}
\renewcommand{\det}{\mathrm{det}}

\newtheorem{proposition}{Proposition}
\allowdisplaybreaks

\title{Likelihood-based inference for the Gompertz model with Poisson errors}
\author{Paolo Onorati\footnote{University of Padua, Italy}, Sofia Ruiz-Suarez\footnote{University of Toronto, Canada}, and Radu V Craiu\footnote{University of Toronto, Canada}}
\date{}

\begin{document}

\maketitle

\begin{abstract}
Population dynamics models play an important role in a number of fields, such as
    actuarial science, demography, and ecology, as they help explain past
    fluctuations and predict future population. The accuracy of these models is
    often influenced by the uncertainty introduced by sampling error.
    Statistical inference for these models can be difficult when, in addition to
    the process' inherent stochasticity, one also needs to account for sampling
    error. Ignoring the latter can lead to biases in the estimation, which in
    turn can produce erroneous conclusions about the system's behavior. The
    Gompertz model is widely used to infer population size dynamics, but a full
    likelihood approach can be computationally prohibitive when sampling error
    is accounted for. We close this gap by developing efficient computational
    tools for statistical inference in the Gompertz model with Poisson sampling
    error based on the full likelihood. The approach is illustrated in both the
    Bayesian and frequentist paradigms. Performance is illustrated with
    simulations and data analysis.
\end{abstract}

\section{Introduction}

Population dynamics models are crucial in both applied and theoretical
mathematical biology, because they help to explain past population fluctuations
and project future population abundances \citep{newman_book}. These models are
used to manage the conservation of species, understand the dynamics of
biological invasions, and assess the response of certain species to changes in
their environment which can range from those produced by human developments to
those triggered by climate change. The accuracy and reliability of these models
are often affected by two major components of uncertainty due to: i) the
process' stochasticity connected to, say, demographic and environmental factors
and ii) the sampling error. 

Ecologists have long recognized the importance of separating observation from
process error in ecological modeling, and significant progress has been achieved
through the use of state-space models to analyze time series of population
fluctuations \citep{Dennis_2006,Hostetler_2015,Methe_2021}. However, including
both process noise and observation error in the model remains challenging, as it
produces significant computational difficulties. To illustrate,
\cite{staples2004estimating} note that the sampling error adds to the
variability in the data leading to positively biased estimators of process
variation and propose a restricted maximum likelihood estimation of the latter
and of the error variance. Subsequently, \citep{linden2009estimating} show
empirically that observation errors can change the autocorrelation structure of
a time series with potential biases resulting from models that ignore such
errors.  

The Gompertz model \citep{gompertz1825xxiv} is widely used to describe and
characterize population dynamics. Its use permeates multiple disciplines,
including actuarial science \citep[e.g.,][]{butt2004application,
baione2018pricing}, demography \citep[e.g.,][]{alexander2017flexible,
tai2018models}, as well as in life sciences where it can be used to model the
growth patterns of animals, plants, tumors and the volume of bacteria
\citep{Winsor1932-ii,tjovre_2017}. Most importantly, the Gompertz model can be
modified to account for sampling variability and, in some clearly structured
problems, it is possible to explicitly formulate the likelihood function in the
presence of sampling error. For example, if the observation noise is assumed to
follow a log-normal distribution, it is feasible to write the likelihood
function \citep{staples2004estimating} and the model can be transformed into a
linear Gaussian state-space model which can be fitted using the Kalman filter
\citep{Dennis_2006}. However, if the sampling error is not log-normal, the
expression of the likelihood becomes more complex. Relevant for this paper is
the study of \cite{Lele} which shows that likelihood-based inference for a
stationary Gompertz model with Poisson sampling errors requires the calculation
of a high-dimensional integral that brings on a prohibitive computational
burden. He proposes a composite likelihood approach that reduces the dimension
of the integral and makes the computation manageable for longer time series.
However, replacing the full likelihood with a composite one comes with an
inferential cost because the resulting confidence, or credible, intervals will
not have nominal coverage. Subsequently, \citet{lele2007data} employed a data
augmentation strategy in both the Gompertz model with normal errors and the
Ricker model with Poisson errors. They used a generic Metropolis-Hastings
algorithm with a random walk proposal implemented through the WinBUGS library to
jointly sample from the posterior distribution of the parameters and latent
variables.

The main contribution of this paper is the construction of computational
algorithms that allow for a full likelihood-based analysis of the Gompertz model
with Poisson errors. Central to our approach is an efficient algorithm for
computing the likelihood, which allows us to conduct both frequentist and
Bayesian inference. The developed methods are available in a new R package
published on GitHub. The paper is organized as follows. In Section
\ref{sec:model} we introduce the notation and the statistical model. Section
\ref{sec:freq} contains the estimation procedure and computational algorithm for
the frequentist approach, which is based on a simulation-aided EM algorithm
\citep{dempster1977,wei1990monte}. The Bayesian analysis relies on a new MCMC
sampler which is described in Section \ref{sec:Bay}. The paper continues with a
Section \ref{sec:exp} of numerical experiments that includes simulations and a
data analysis. A discussion of open questions and future research directions
closes the paper in Section \ref{sec:disc}.

\section{Statistical Model}
\label{sec:model}

Consider a population observed at times $t=1,2,\ldots, T$. The Gompertz model
with stochastic errors establishes a probabilistic model for the evolution of
$N_t$, the population size at time $t$, for $1\le t \le T$. Specifically, assume
that for all $1\le t \le T-1$,
\begin{equation}
	N_{t+1}=N_t\exp(a+b\log(N_t)+\varepsilon_{t+1}),
	\label{gomp1}
\end{equation}
where $a$ is an individual intrinsic growth rate parameter, $b$ links the
current measure with the past one, and $\varepsilon_t$ are
independent errors normally distributed: $\varepsilon_t \sim N(0, \sigma^2)$.
The logarithmic transformation of $N_t$ in \eqref{gomp1} leads to
\begin{equation} \label{eq:logGomp}
    Z_{t+1}=a+(1+b)Z_t+\varepsilon_{t+1}
\end{equation}
where $Z_t=\log(N_{t})$. Equation \eqref{eq:logGomp} implies that the Gompertz
model corresponds, on the logarithmic scale, to an autoregressive process of
order $1$. If we assume that $|1+b|<1$, then for any subset of $n$ observations,
the $n$-dimensional marginal distribution of the log transformed population
size, $Z_t$, has a multivariate normal distribution
$(Z_{i_1},Z_{i_2},\dots,Z_{i_n})^\prime \sim N_n(\bm\mu, \Sigma)$ where
\begin{equation} \label{eq:n-nomalMarginal}
    \bm\mu=(\theta_1,\theta_1,\dots,\theta_1)^\prime \, , \,
    \Sigma = \theta_2 B \, ,
    \text{ and } 
    B_{jk} = (1+b)^{ \vert i_j-i_k \vert} 
\end{equation}
and $\theta_1 = -a/b$ and $\theta_2 = -\sigma^2 / \big[ b(2+b) \big]$. 
In what follows, we denote $\bsTheta=(\theta_1,\theta_2,b)$.

In this paper, we consider the case where the exact population size $N_t$ is
unknown. Instead, we assume that at time $t\in\{1,\dots,T\}$, $N_t^*$ is the
observed population size, whose conditional probability mass function is denoted
$\pi(N_t^*|N_t,\phi)$ and is indexed by the parameter $\phi$. Furthermore, we
assume that the random variables $\{N^*_t| \; 1\le t \le T\}$ are independent
conditionally on $\{N_t: \; 1\le t \le T\}$.  The observed-data likelihood is
obtained by taking advantage of the conditional distribution of $N_t^*|N_t$ and
averaging the missing true population values $\{N_1,\ldots,N_T\}$.
\begin{equation}
  \label{eq:likelihood}
  L(\bsTheta, \phi \vert \boldsymbol{N}^\ast) = \int \int \dots \int \pi(N_1^*|N_1,\phi) \pi(N_2^*|N_2,\phi) \dots \pi(N_T^*|N_T,\phi) \pi(\bsN\vert \bsTheta) \, dN_1 \, dN_2 \dots \, dN_T \, , \\
\end{equation}
where $\pi(\bsN \vert \bsTheta)$ indicates the joint distribution of the
unobserved time series of exact population sizes.

If the sampling error distribution $\pi(N_t^*|N_t,\phi)$ is assumed to be
log-normal, then it is possible to explicitly formulate the complete likelihood
function as in \cite{Dennis_2006}. However, for other sampling error
distributions, the likelihood \eqref{eq:likelihood} cannot be expressed in
closed form using elementary functions. Here we focus on the case of a Poisson
sampling error distribution, i.e. $N_t^* \sim \mbox{Poisson}(N_t)$, for all
$1\le t\le T$. In this case, no additional parameters, $\phi$, are associated
with the sampling error distribution.  However, the use of a Poisson
distribution leads to computational challenges because the likelihood is not
available in closed form, as discussed by \citep{Lele}. In
Section~\ref{sec:freq} we present a method for directly addressing the
expression \eqref{eq:likelihood} to perform parameter inference. Our approach is
very general and can also be implemented in a Bayesian framework, as we show in
Section~\ref{sec:Bay}. The performance of the proposed algorithms is examined in
Section \ref{sec:simStudy}.

\section{Maximum likelihood based inference}
\label{sec:freq}

We describe a frequentist inferential method for the Gompertz model with Poisson
observation errors. In particular, we are interested in computing the maximum
likelihood estimator (MLE) for $\bsTheta$ and producing confidence intervals.
The proposed algorithm takes advantage of the data augmentation strategy implied
by \eqref{eq:likelihood}. It is apparent that if the population sizes $\bsN$
were available, then the inference would be straightforward, since it would rely
on the analysis of a first-order autoregressive model for $Z_t = \log N_t$.
However, since population sizes $N_t, t \in \{1,2,...,T\}$ are unobserved latent
variables, we adopt a data augmentation approach.

The expectation-maximization (EM) algorithm is a powerful iterative procedure
that is used to estimate the parameters of models with missing data. In our
case, $\bsN^*$ is the vector of observed data, $\bsZ = \log \bsN$ is the vector
of missing data, and $\bsTheta^{(k)}$ is the value of the parameter estimate in
the $k$-th iteration. The EM algorithm \citep{dempster1977} iteratively computes
$\bsTheta^{(k + 1)}=\arg \max_{\bsTheta} Q(\bsTheta|\bsTheta^{(k)})$ where the
$Q$ function is
\begin{equation*}
   Q(\bsTheta|\bsTheta^{(k)}) =  \E [\log \pi (\bsN^*, \bsZ \vert \bsTheta)\vert \bsN^*,  \bsTheta^{(k)}] \, .
\end{equation*}
The calculation of the conditional expectation of the complete data
log-likelihood, $\log \pi(\bsN^*, \bsZ \vert \bsTheta)$, represents the E-step
and  its optimization as a function of $\bsTheta$ is the M-step. The algorithm
cycles between the E- and M-steps until the change in $\bsTheta^{(k)}$ becomes
negligible. 

In the specific case of the Gompertz model with Poisson sampling error, the
complete data log-likelihood can be separated as
\begin{equation*}
    \log \pi (\bsN^*, \bsZ \vert \bsTheta ) = \log \pi (\bsN^* \vert \bsZ) + \log \pi (\bsZ \vert \bsTheta ) \, ,
\end{equation*}
and only the second term depends on $\bsTheta$. Nonetheless, we cannot compute
the expectation required in the E-step in closed form, so  a Monte Carlo
strategy must be used. Specifically, we approximate the expectation with a Monte
Carlo average
\begin{equation}
    \E [\log \pi (\bsZ \vert \bsTheta ) \vert \bsN^*, \bsTheta = \bsTheta^{(k)}] \approx \frac{1}{J} \sum_{j = 1}^J \log \pi(\bstilZ_j \vert \bsTheta) \, ,
    \label{mcem}
\end{equation}
where $\bstilZ_j$ are sampled from $\bsZ \vert \bsN^*, \bsTheta = \bsTheta^{k}$.
Replacing the expectation with a Monte Carlo estimate is the idea behind the
so-called Monte Carlo EM (MCEM) introduced by \cite{wei1990monte} and further
expanded by \cite{mcculloch1994maximum, chan1995monte,booth1999maximizing}, and
\cite{caffo2005}. 

We start the MCEM algorithm with a value of $J = 10^3$ in \eqref{mcem} and,  as
prescribed by \cite{booth1999maximizing} and \cite{caffo2005}, we increase the
value of $J$ with the number of iterations in order to achieve higher precision
in the estimate of the expected complete data log-likelihood. Specifically, we
use the method proposed in \citet{caffo2005} to determine whether an increase in
$J$ is necessary or not.  If so, we double the current value of $J$ up to a
maximum value of $2 \times 10^4$. The conditional distribution $\bsZ \vert
\bsN^*, \bsTheta = \bsTheta^{(k)}$ is nonstandard, so we must customize a Gibbs
sampler to obtain the necessary draws. In the following, we omit the superscript
$^{(k)}$ for ease of notation.

According to the Gibbs sampler design, we need to obtain samples from the full
conditional distributions of each latent variable, given the observed data, the
current parameter values, and the remaining unobserved population sizes. For
brevity, we denote $\bsZ_{-t}$ the vector $\bsZ$ from which the $t$ -th
component, $Z_t$ is excluded. Thus, we need to sample from
\begin{equation} \label{eq:full_conditionals}
    Z_t \vert \bsZ_{-t}, \bsN^\ast, \bsTheta = 
        \left\{ \begin{array}{lcc}
            Z_1 \vert Z_2, N^\ast_1, \bsTheta  & \text{if} & t=1 \\
            \\ Z_t \vert Z_{t-1}, Z_{t+1}, N^\ast_t, \bsTheta  & \text{if} & 1 < t < T \\
            \\ Z_T \vert Z_{T-1}, N^\ast_T, \bsTheta  & \text{if} & t=T \\
        \end{array} \right. \, .    
\end{equation}
The expression above exploits the independence of the components of $\bsN^\ast$
conditionally on $\bsZ$ and the sparse structure of the inverse of the
correlation matrix $B$. These facts dramatically improve the mixing
of the Gibbs sampler.

From expression \eqref{eq:full_conditionals}, we obtain the unnormalized
density for every full conditional distribution of $Z_t$, that is,
\begin{equation} \label{eq:full_conditional_densities}
    \pi(Z_t \vert \bsZ_{-t}, \bsN^\ast, \bsTheta) \propto 
    \left\{ \begin{array}{lcc}
        \pi(N^\ast_1 \vert Z_1) \, \pi(Z_1 \vert Z_2, \bsTheta)  & \text{if} & t=1 \\
        \\ \pi(N^\ast_t \vert Z_t) \, \pi(Z_t \vert Z_{t-1}, Z_{t+1}, \bsTheta)  & \text{if} & 1 < t < T \\
        \\ \pi(N^\ast_T \vert Z_T) \, \pi(Z_T \vert Z_{T-1}, \bsTheta)  & \text{if} & t=T \\
    \end{array} \right. \, ,  
\end{equation}
and the second factor in the above expression is always a Gaussian density with
mean $\mu_t$ and variance $\tau^2_t$ given by
\begin{align} \label{eq:full_conditional_meanvar}
    \mu_t = 
    \left\{ \begin{array}{lcc}
        \frac{\theta_1 \sigma^2 + (Z_2 - a)(1 + b)\theta_2}{\sigma^2 + (1 + b)\theta_2}  & \text{if} & t=1 \\
        \\ \frac{a + (1 + b)(Z_{t+1} + Z_{t-1} - a)}{1 + (1 + b)^2}  & \text{if} & 1 < t < T \\
        \\ a + (1 + b)Z_{T-1}  & \text{if} & t=T \\
    \end{array} \right. , \quad
    \tau^2_t = 
    \left\{ \begin{array}{lcc}
        \frac{\sigma^2 \theta_2}{\sigma^2 + (1 + b)\theta_2}  & \text{if} & t=1 \\
        \\ \frac{\sigma^2}{1 + (1 + b)^2}  & \text{if} & 1 < t < T \\
        \\ \sigma^2  & \text{if} & t=T \\
    \end{array} \right. \, ,
\end{align}
Therefore, expression \eqref{eq:full_conditional_densities} is equal to the
product between a Gaussian density and a Poisson mass probability, so that 
\begin{equation}
    \label{eq:pogau}
    \pi(Z_t \vert \bsZ_{-t}, \bsN^\ast, \bsTheta) \propto \exp \left( -e^{Z_t} - \frac{(Z_t - \mu_t)^2}{2 \tau^2_t} \right) \, ,
\end{equation}
where $\mu_t$ and $\tau^2_t$ are given by \eqref{eq:full_conditional_meanvar}
and channel the dependence between $Z_t$ and $\bsZ_{-t}$. 

We sample from the above density  using an accept-reject algorithm with a
$N(\xi, \omega^2)$ serving as the proposal density. The following proposition
provides the upper bound between the target and the proposal and, in the special
case of $\omega^2 = \tau^2_t$, the best value for $\xi$.
\begin{proposition}
  Using the above notation the following hold:

  i) The ratio between expression \eqref{eq:pogau} and the density of $N(\xi,
    \omega^2)$ is unbounded if $\omega^2 < \tau^2_t$.
    
  ii) If $\omega^2 \ge \tau^2_t$ then there exits an unique maximum equal to
   	\begin{equation*}
      \widehat{Z}_t = \left\{
				\begin{array}{lcc}
      		\log \left( \frac{N^\ast_t \tau^2_t - \xi + \mu_t}{\tau^2_t} \right) & \text{if} & \omega^2 = \tau^2_t
       		\\ \\
       		\frac{N^\ast_t \omega^2 \tau^2_t - \xi \tau^2_t + \mu \omega^2}{\omega^2 - \tau^2_t} \W_0 \left( \frac{\omega^2 \tau^2}{\omega^2 - \tau^2_t} \exp \left( \frac{N^\ast_t \omega^2 \tau^2_t - \xi \tau^2_t + \mu \omega^2}{\omega^2 - \tau^2_t} \right) \right) & \text{if} & \omega^2 > \tau^2_t \end{array} \right. \, ,
   	\end{equation*}
   	where $\W_0$ denotes the upper branch of the Lambert $\W$ function. 

  iii) When $\omega^2 = \tau^2_t$ the value of $\xi$ that minimizes the maximum
  $\widehat{Z}_t$ is
    \begin{equation*}
        \hat \xi = N^\ast_t \tau^2_t + \mu_t - \W_0 \left( \tau^2_t \exp(N^\ast_t \tau^2_t + \mu_t) \right) \, .
    \end{equation*}
\end{proposition}
\begin{proof}

  i) The ratio between the target and the proposal density is proportional to
    \begin{equation}
      h(Z_t) = \exp \left( Z_t N^\ast_t -e^{Z_t} - \frac{(Z_t - \mu_t)^2}{2 \tau^2_t} + \frac{(Z_t - \xi)^2}{2 \omega^2} \right) \, .
      \label{AR-ratio}
    \end{equation}
    It is easy to show that is unbounded if $\omega^2 < \tau^2_t$. 
    
  ii) We compute the first and second log-derivative and obtain
    \begin{align*}
        \frac{\d \log h(Z_t)}{\d Z_t} &= N^\ast_t + \frac{Z_t - \xi}{\omega^2} - \frac{Z_t - \mu_t}{\tau^2_t} - e^{Z_t} \, , \\
        \frac{\d^2 \log h(Z_t)}{\d Z_t ^ 2} &= \frac{1}{\omega^2} - \frac{1}{\tau^2_t} - e^{Z_t} \, .
    \end{align*}
    Therefore, the second derivative is always negative as long as $\omega^2 \ge
    \tau^2_t$; in this case  $h(\cdot)$ is log-concave and the unique maximum
    occurs at the value $Z_t=\widehat{Z}_t$ where the first derivative is equal
    to $0$. When $\omega^2 \ge \tau^2_t$, setting the first derivative to $0$
    yields:
    \begin{equation*}
      e^{\widehat{Z}_t} = \frac{N^\ast_t \omega^2 \tau_t^2 - \xi \tau^2_t + \mu_t \omega^2}{\omega^2 \tau^2_t} - \widehat{Z}_t \, \frac{\omega^2 - \tau_t^2}{\omega^2 \tau^2_t} \, .
    \end{equation*}
    If we settle on $\omega^2 > \tau^2_t$, after some algebra we obtain
    \begin{equation*}
      \widehat{Z}_t = \frac{N^\ast_t \omega^2 \tau^2_t - \xi \tau^2_t + \mu \omega^2}{\omega^2 - \tau^2_t} \W_0 \left( \frac{\omega^2 \tau^2}{\omega^2 - \tau^2_t} \exp \left( \frac{N^\ast_t \omega^2 \tau^2_t - \xi \tau^2_t + \mu \omega^2}{\omega^2 - \tau^2_t} \right) \right) \, ,
    \end{equation*}
    where $\W_0$ denotes the upper branch of the Lambert $\W$ function.
    However, this solution is cumbersome to work with and, more importantly, 
    does not guarantee a more efficient solution.
    
    iii) If we set $\omega^2 = \tau^2_t$ we obtain 
    \begin{equation*}
      \widehat{Z}_t = \log \left( \frac{N^\ast_t \tau^2_t - \xi + \mu_t}{\tau^2_t} \right) \, .
    \end{equation*}
    Replacing $\omega^2=\tau_t^2$ in \eqref{AR-ratio} we obtain  
    \begin{equation*}
      \log h( \widehat{Z}_t) = \widehat{Z}_t N^\ast_t + \frac{(\widehat{Z}_t -\xi)^2 - (\widehat{Z}_t -\mu_t)^2}{2 \tau^2_t} - e^{\widehat{Z}_t} \, .
    \end{equation*}
    The value $\hat\xi$ which minimizes $\log h( \widehat{Z}_t)$ is found using
    \begin{align*}
      \frac{\d \log h(\widehat{Z}_t)}{\d \xi} &= \frac{\d \log h(x)}{\d x} \Bigg\vert_{x = \widehat{Z}_t} \frac{\d \widehat{Z}_t}{\d \xi} + \frac{\d \log h(x)}{\d \xi} \Bigg\vert_{x = \widehat{Z}_t}
			\\
      &= \frac{\d \log h(x)}{\d \xi} \Bigg\vert_{x = \widehat{Z}_t}
			\\
      &=-\frac{\widehat{Z}_t - \xi}{\tau^2_t} \, .
    \end{align*}
    Since the second derivative is always positive, $\log h(\widehat{Z}_t)$ is a log-convex function with an unique minimum at $\xi = \widehat{Z}_t$. However, since
    $\widehat{Z}_t$ is itself a function of $\xi$, from the following condition
    \begin{equation*}
      \xi = \log \left( \frac{N^\ast_t \tau^2_t - \xi + \mu_t }{\tau^2_t} \right) \, ,
    \end{equation*}
    after some algebra, we obtain the final expression 
    \begin{equation*}
      \hat \xi = N^\ast_t \tau^2_t + \mu_t - \W_0 \left( \tau^2_t \exp(N^\ast_t \tau^2_t + \mu_t) \right) \, .
    \end{equation*}
\end{proof}

The proposed algorithm falls within the class of Markov chain Monte Carlo EM
(MCMC-EM) since it is an MCEM optimization algorithm in which  the draws
required to complete the E-step are obtained using the Gibbs sampler. For the
efficiency of the algorithm, it is important to choose carefully the
initialization point. Regarding the starting point $\bsTheta^{(0)}$, we set it
equal to the method of moments estimator that is provided by the following
proposition.
\begin{proposition}
  In the Gompertz model with Poisson sampling error distribution, the mean,
  variance and covariances for $\{N_t^*, 1\le t \le T\}$ are:
  \begin{align*}
    \mbox{i)} \; \E(N^*_t) &= \exp \Big(\theta_1 + \frac{\theta_2}{2} \Big) \, ,
     \\
    \mbox{ii)} \; \Var(N^*_t) &= \exp \big(2\theta_1 + \theta_2 \big) \Big(\exp \big(\theta_2 \big) - 1 \Big) \, ,
     \\
     \mbox{iii)} \; \Cov(N_t^*, N_{t + h}^*) &= \exp \big(2\theta_1 + \theta_2 \big) \Big( \exp \big(\theta_2 (1 + b)^h \big) - 1 \Big) \, .
  \end{align*}
\end{proposition}
\begin{proof}
  Equation \eqref{eq:n-nomalMarginal} implies
  \begin{align*}
    \E(Z_t) = \theta_1 \, , \, \Var(Z_t) = \theta_2 \, , \, \Cov(Z_t, Z_{t + h}) = \theta_2 (1 + b)^{|h|} \, .
  \end{align*}
  On the other hand, $N_t = \exp(Z_t)$ and based on the from the expression for
  the moment generating function of a Gaussian random variable, we get
  \begin{equation*}
    \E \big(\exp(c Z_t) \big) = \exp \Big(c \theta_1 + \frac{c^2 \theta_2}{2} \Big) \, ,    
  \end{equation*}
  and thus the following expressions hold
  \begin{align*}
    \E(N_t) &= \E \big(\exp(Z_t) \big) = \exp \Big(\theta_1 + \frac{\theta_2}{2} \Big) \, ,
    \\
    \Var(N_t) &= \E \big(\exp(2 Z_t) \big) - \E^2 \big(\exp(Z_t) \big) = \exp(2 \theta_1 + \theta_2) (\exp(\theta_2) - 1) \, ,
    \\
    \Cov(N_t, N_{t + h}) &= \E \big(\exp(Z_t + Z_{t + h}) \big) - \E^2 \big(\exp(Z_t) \big) = \exp \big(2 \theta_1 + \theta_2 \big) \big( \exp(\theta_2 (1 + b)^h) - 1 \big) \, .
  \end{align*}
  The last expression is valid because $Z_t + Z_{t + h} \sim N \big(2 \theta_1,
  2\theta_2 (1 + (1+b)^h) \big)$ and $\E \big(\exp(Z_t) \big) = \E
  \big(\exp(Z_{t + h}) \big)$. Furthermore, since $N^*_t \vert N_t
  \overset{ind}{\sim} \Po(N_t)$ we have $\E(N^*_t \vert N_t) = \Var(N^*_t \vert
  N_t) = N_t$ and $\Cov(N^*_t, N^*_{t + h} \vert N_t, N_{t + h}) = 0$ due to
  conditional independence. Using the law of total expectation, the law of total
  variance and the law of total covariance is easy to obtain the first two
  moments for $N^*_t$, as
  \begin{align*}
    \E(N^*_t) &= \E \big(\E(N^*_t \vert N_t) \big) = \exp \Big(\theta_1 + \frac{\theta_2}{2} \Big) \, ,
    \\
    \Var(N^*_t) &= \E \big(\Var(N^*_t \vert N_t) \big) + \Var \big(\E(N^*_t \vert N_t) \big) = \exp \big(2\theta_1 + \theta_2 \big) \Big(\exp \big(\theta_2 \big) - 1 \Big) \, ,
    \\
    \Cov(N_t^*, N_{t + h}^*) &= \E \big(\Cov(N^*_t, N^*_{t + h} \vert N_t, N_{t + h}) \big) + \Cov \big(\E(N^*_t \vert N_t), \E(N^*_{t + h} \vert N_{t + h}) \big) 
    \\
    &= \exp \big(2\theta_1 + \theta_2 \big) \Big( \exp \big(\theta_2 (1 + b)^h \big) - 1 \Big) \, .
  \end{align*}
\end{proof}
We set the starting point $\bsTheta^{(0)}$ in order to match the first two
moments provided by the above proposition with their sample counterparts.
Regarding the covariance, we set $h = 1$ because the first lag is the most
efficient to estimate.

We use a sampling importance resampling strategy for the initialization of
$\bsZ$. The importance density is equal to
\begin{align*}
  q(\bsZ) &= \pi \big(Z_1 \vert N^*_1, Z_2 = \log(N^*_2) \big) \pi \big(Z_2 \vert N^*_2, Z_1, Z_3 = \log(N^*_3) \big) \dots
  \\
  &\times \pi \big(Z_{t - 1} \vert N^*_{t - 1}, Z_{t - 2}, Z_t = \log(N^*_t) \big) \pi \big(Z_t \vert N^*_t, Z_{t - 1} \big) \, .
\end{align*}
Thus, the importance density for $Z_t$ is equal to its full conditional used in
the Gibbs sampler; the full conditional is obtained setting $Z_{t + 1} =
\log(N_{t + 1}^*)$. Therefore, we draw $10,000$ values of $\bsZ$ from $q(\cdot)$
and compute the importance weights. The initial value is then selected with
probability proportional to these weights.

The proposed MCMC-EM method can be used computing the asymptotic variance of the
MLE too. As noted in \citet{caffo2005}, it is easy to use the output of the
sampling algorithm to calculate the inverse of the Fisher information matrix
using the method of \citet{louis1982}. This allows us to produce asymptotic
variance estimates for $\hat \bsTheta$.

\section{Bayesian Inference}
\label{sec:Bay}

The Bayesian paradigm offers a different probabilistic mechanism to estimate
finite sample variances for the estimators of interest, allows principled ways
to incorporate prior knowledge when it is available, and to integrate model
uncertainty into the predictions via model averaging.  The crux of the approach
is the posterior distribution which encodes all the uncertainty after observing
the data. In the Gompertz model with Poisson errors, the posterior distribution
is analytically intractable, so we must study it using MCMC sampling. The data
augmentation strategy presented in Section~\ref{sec:freq} also plays a central
role in the design of the MCMC algorithm. Perhaps surprisingly, the numerical
experiments show that the algorithm for sampling the Bayesian posterior is much
more efficient than the MCMC-EM in terms of computation time.

To perform a fair comparison with the MLE, we propose using a weakly informative
prior. We use an uniform prior for $b$, that is, $b \sim U(-2,0)$, and
normal-inverse gamma priors for $\theta_1$ and $\theta_2$,
\begin{equation*}
    \theta_2 \sim \igamma(\varphi_1, \varphi_2) \text{ and } \theta_1 \vert \theta_2 \sim N(\eta_1, \eta_2 \theta_2) \, .
\end{equation*}
We also assume that $\theta_1,\theta_2$ are a priori independent of $b$. We
choose values for the hyperparameters $\varphi_1, \varphi_2, \eta_1$, and
$\eta_2$ that lead to a weakly informative prior. We set $\varphi_1 = \varphi_2
= 0.1$, $\eta_1 = 0$, and $\eta_2 = 100$. The sampling algorithm's steps do not
depend on the particular values we choose for the hyperparameters, so we
describe them in terms of generic values. 

The dependent draws from the conditional distribution of all parameters and
augmented data $\pi(\theta_1, \theta_2, b, \bsZ, \vert N^\ast)$ are obtained
using a Gibbs sampler.

Let $\theta_1^{(k)}, \theta_2^{(k)}, b^{(k)}, \bsZ^{(k)}$ be the sample values
in the $k$-th iteration of the MCMC algorithm. The starting values
$\theta_1^{(0)}, \theta_2^{(0)}, b^{(0)}$ are set equal to the values produced
by the moment estimator method and $\bsZ^{(0)}$ is drawn using sampling
importance resampling, as in the MCMC-EM initialization.

Thus, given $\theta_1^{(k)}, \theta_2^{(k)}, b^{(k)}$, and $\bsZ^{(k)}$, we
obtain $\theta_1^{(k+1)}, \theta_2^{(k+1)}, b^{(k+1)}$, and $\bsZ^{(k+1)}$ using
the following update scheme:
\begin{enumerate}
  \item \label{cond_Zt} For $t = 1, 2, \dots, T$, sample $Z_t^{(k+1)} \vert N^\ast_t,
  \theta_1^{(k)}, \theta_2^{(k)}, b^{(k)}, \bsZ_{-t}^{(k)}$.
	\item Sample $\theta_1^{(k+1)} ,\theta_2^{(k+1)}, b^{(k+1)} \vert
  \bsZ^{(k+1)}$. 
  \begin{enumerate}
      \item \label{cond_b} Sample $b^{(k+1)} \vert \bsZ^{(k+1)}$.
      \item \label{cond_theta2} Sample $\theta_2^{(k+1)} \vert b^{(k+1)}, \bsZ^{(k+1)}$.
      \item Sample \label{cond_theta1} $\theta_1^{(k+1)} \vert \theta_2^{(k+1)}, b^{(k+1)}. \bsZ^{(k+1)}$.
  \end{enumerate}    
\end{enumerate}
In step (\ref{cond_Zt}), we use the acceptance-rejection algorithm described in
section \ref{sec:freq}. In steps (\ref{cond_theta2}) and (\ref{cond_theta1}) the
full conditionals are standard, since
\begin{align*}
    \theta_2 \vert b, \bsZ &\sim  \igamma \left( \phi_1 + \frac{T}{2}, \phi_2 + \frac{1}{2} (\bsZ - \eta_1 1_T)^\prime (\eta_2 1_T 1^\prime_T + B)^{-1} (\bsZ - \eta_1 1_T) \right) \, ,
    \\
    \theta_1 \vert \theta_2, b, \bsZ &\sim N \left( \frac{\eta_1 + \eta_2 1^\prime_T B^{-1} \bsZ}{1 + \eta_2 1^\prime_T B^{-1} 1_T}, \frac{\eta_2 \theta_2}{1 + \eta_2 1^\prime_T B^{-1} 1_T} \right) \, ,
\end{align*}
where $1_T$ denotes the $T$-dimensional vector of ones. In contrast, step
(\ref{cond_b}) is not standard. To sample from the conditional density of $b$
given $\bsZ$ we use another accept-reject algorithm. The proposal is the uniform
prior itself; thus the upper bound between the target density and the proposal
density is the maximum of $\pi(\bsZ \vert b)$. Since the prior of $(\theta_1,
\theta_2)$ is a normal-inverse gamma distribution, straightforward calculation
yields $Z \vert \theta_2, b \sim N_T \big(\eta_1 1_T, \theta_2 (\eta_2 1_T
1^\prime_T + B) \big)$ and then
\begin{align}
    \nonumber
    \pi&(\bsZ \vert b) = \int_{\theta_2} \int_{\theta_1} \pi(\bsZ \vert \theta_1, \theta_2, b) \pi(\theta_1, \theta_2) \d \theta_1 \d \theta_2
    \\
    \nonumber
    &\propto \det^{-\frac{1}{2}}(\eta_2 1_T 1^\prime_T + B) \int_0^{+\infty} \theta^{-\frac{2 \phi_1 + T}{2} - 1}_2 \exp \left(-\frac{1}{\theta_2} \Big(\phi_2 + \frac{1}{2} (\bsZ - \eta_1 1_T)^\prime (\eta_2 1_T 1^\prime_T + B)^{-1} (\bsZ - \eta_1 1_T) \Big) \right) \d \theta_2
    \\
    \label{eq:PDF_b}
    &\propto \det^{-\frac{1}{2}}(\eta_2 1_T 1^\prime_T + B) \left(1 + \frac{1}{2 \phi_2}(\bsZ - \eta_1 1_T)^\prime (\eta_2 1_T 1^\prime_T + B)^{-1} (\bsZ - \eta_1 1_T) \right)^{-\frac{2 \phi_1 + T}{2}} \, .
\end{align}
The expression above can be further simplified using the matrix determinant
lemma, Sherman-Morrison formula, and the closed-form expression for $B^{-1}$.
The latter is available because $B$ is the correlation matrix of a stationary
autoregressive process of order $1$.
\begin{proposition}
    In the Gompertz model with Poisson sampling error distribution, the density
    of $\bsZ \vert b$ is
    \begin{equation}
        \label{eq:PDF_b_simpl}
        \begin{split}
            \pi(\bsZ \vert b) &\propto \big(1 - r^2 \big)^{1 - \frac{T}{2}} \Big( \big(\eta_2 (T - 2) - 1 \big)r^2 - 2 \eta_2 (T - 1) r + \eta_2 T + 1 \Big)^{-\frac{1}{2}} \\
            & \times \Bigg(1 + \frac{r^2 \sum\limits_{s=2}^{T-1} W^2_s - 2 r \sum\limits_{h=1}^{T-1} W_h W_{h+1} + \sum\limits_{t=1}^{T}W^2_t}{2 \, \phi_2 \, (1 - r^2)} - \frac{\eta_2 (1-r^2)}{\big(\eta_2 (T - 2) - 1 \big)r^2 - 2 \eta_2 (T - 1) r + \eta_2 T + 1} \\
            & \times \frac{r^2 \Big(\sum\limits_{s=2}^{T-1} W_s \Big)^2 - 2 r \sum\limits_{s=2}^{T-2} W_s \sum\limits_{t=1}^{T} W_t + \Big(\sum\limits_{t=1}^{T} W_t \Big)^2}{2 \, \phi_2 \, (1 + r^2)} \Bigg)^{-\phi_1 - \frac{T}{2}} \, ,
        \end{split}
    \end{equation}
    where $W_t = Z_t - \eta_1$ and $r = 1 + b$.
\end{proposition}
\begin{proof}
    Let $r = 1 + b$ and $W = Z - \eta_1 1_T$. Then
    \citep{hamilton1994time}[Section 5.2]
    \begin{align*}
        B &= \begin{bmatrix}
        1         & r         & r^2       & \dots     & r^{T - 1} \\
        r         & 1         & r         & \dots     & r^{T - 1} \\
        \dots     & \dots     & \dots     & \dots     & \dots     \\
        r^{T - 1} & r^{T - 2} & r^{T - 3} & r^{T - 4} & 1         
        \end{bmatrix} \, , \, \det(B) = (1 - r^2)^{T - 1} \, ,
        \\
        B^{-1} &= \frac{1}{1-r^2}
        \begin{bmatrix}
        1     & -r     &        &        &    \\
        -r    & 1+r^2  & -r     &        &    \\
              & \ddots & \ddots & \ddots &    \\
              &        & -r     & 1+r^2  & -r \\
        0     &        &        & -r     & 1  
        \end{bmatrix}.
    \end{align*}
    Using the matrix determinant lemma and Sherman-Morrison formula, it is
    obtained
    \begin{align*}
        \det(\eta_2 1_T 1^\prime_T + B) &= \det(B) (1 + \eta_2 1^\prime_T B^{-1} 1_T) \, ,
        \\
        (\eta_2 1_T 1^\prime_T + B)^{-1} &= B^{-1} - \eta_2 \frac{B^{-1} 1_T 1^\prime_T B^{-1}}{1 + \eta_2 1^\prime_T B^{-1} 1_T} \, .
    \end{align*}
    This implies
    \begin{equation}
        \label{eq:PDF_b_quasi_simpl}
        \pi(Z \vert b) \propto \det^{-\frac{1}{2}}(B) (1 + \eta_2 1^\prime_T B^{-1} 1_T)^{-\frac{1}{2}} \Big(1 + \frac{W^\prime B^{-1} W}{2 \phi_2} - \frac{\eta_2}{(1 + \eta_2 1^\prime_T B^{-1} 1_T)} \frac{W^\prime B^{-1} 1_T 1^\prime_T B^{-1} W}{2 \phi_2} \Big) \, .
    \end{equation}
    Consider separately $1 + \eta_2 1^\prime_T B^{-1} 1_T$. It is
    straightforward that $1^\prime_T B^{-1} 1_T$ is the sum of all components of
    the matrix $B^{-1}$; in this matrix there are $(T - 2)$, $2 (T - 1)$, and
    two times $1 + r^2$, $-r$, and one respectively. Then
    \begin{align}
        \nonumber
        1 + \eta_2 1_T^{\top} B^{-1} 1_T &= 1 + \frac{\eta_2}{1-r^2} \left( 2 + (1+r^2)(T-2) - 2r (T-1) \right)
        \\
        \nonumber
        &= \frac{1 - r^2 + \eta_2 \big(r^2 (T - 2) - 2r (T - 1) + T \big)}{1-r^2}
        \\
        \label{eq:PDF_b_1cor}
        &= \frac{\big(\eta_2 (T - 2) - 1 \big)r^2 - 2\eta_2 (T - 1)r + \eta_2 T + 1}{1 - r^2} \, .
    \end{align}
    We now focus our attention on  $W^\prime B^{-1} W$, which can be simplified
    as follows:
    \begin{align}
        \nonumber 
        W^\prime B^{-1} W &= \frac{1}{1-r^2}
        \begin{bmatrix}
        W_1 \\[6pt]
        W_2 \\[6pt]
        \vdots \\[6pt]
        W_T
        \end{bmatrix}^\prime
        \begin{bmatrix}
        1 & -r & 0 & \cdots & 0 \\
        -r & 1+r^2 & -r & \cdots & 0 \\
        0 & -r & 1+r^2 & \ddots & \vdots \\
        \vdots & \vdots & \ddots & \ddots & -r \\
        0 & 0 & \cdots & -r & 1
        \end{bmatrix}
        \begin{bmatrix}
        W_1 \\[6pt]
        W_2 \\[6pt]
        \vdots \\[6pt]
        W_T
        \end{bmatrix} 
        \\[1em]
        &= \frac{1}{1-r^2}
        \begin{bmatrix}
        W_1 \\[20pt]
        W_2 \\[20pt]
        \vdots \\[20pt]
        W_T
        \end{bmatrix}^\prime
        \begin{bmatrix}
        W_1 - r W_2 \\
        (1+r^2)W_2 - r(W_1+W_3) \\
        (1+r^2)W_3 - r(W_2+W_4) \\
        \vdots \\
        (1+r^2)W_t - r(W_{t-1}+W_{t+1}) \\
        \vdots \\[6pt]
        (1+r^2)W_{T-1} - r(W_{T-2}+W_T) \\
        W_T - r W_{T-1}
        \end{bmatrix} \nonumber
        \\[1em]
        \nonumber
        &= \frac{1}{1-r^2}
        \left( W_1^2 - r W_1 W_2 + W_T^2 - r W_{T-1}W_T + \sum_{s=2}^{T-1} \Big[(1 + r^2) W_s^2 - r W_s (W_{s-1}+W_{s+1}) \Big] \right)
        \\
        \nonumber
        &= \frac{1}{1-r^2} \left( \sum_{t=1}^T W_t^2 -r W_1 W_2 - r \sum_{s=2}^{T-1} W_s W_{s+1} - r W_T W_{T_1} - r \sum_{s=2}^{T-1} W_s W_{s-1} + r^2 \sum_{s=2}^{T - 1} W_s^2 \right)
        \\
        \label{eq:PDF_b_2cor}
        &= \frac{1}{1-r^2} \left( r^2 \sum_{s=2}^{T-1} W_s^2 - 2r \sum_{s=2}^{T-1} W_s W_{s+1} + \sum_{t=1}^T W_t^2 \right) \, .
    \end{align}
   Similarly, $W^\prime B^{-1} 1_T$ yields
    \begin{align}
        \nonumber
        W^\prime B^{-1} 1_T &=  \frac{1}{1-r^2}
        \begin{bmatrix}
        W_1 \\[6pt]
        W_2 \\[6pt]
        \vdots \\[6pt]
        W_T
        \end{bmatrix}^\prime
        \begin{bmatrix}
        1 & -r & 0 & \cdots & 0 \\
        -r & 1+r^2 & -r & \cdots & 0 \\
        0 & -r & 1+r^2 & \ddots & \vdots \\
        \vdots & \vdots & \ddots & \ddots & -r \\
        0 & 0 & \cdots & -r & 1
        \end{bmatrix}
        \begin{bmatrix}
        1 \\[6pt]
        1 \\[6pt]
        \vdots \\[6pt]
        1
        \end{bmatrix}
        \\[1em]
        \nonumber
        &= \frac{1}{1-r^2}
        \begin{bmatrix}
        W_1 \\[3pt]
        W_2 \\[3pt]
        \vdots \\[3pt]
        W_T
        \end{bmatrix}^\prime
        \begin{bmatrix}
        1 - r \\
        r^2 -2r + 1 \\
        \dots \\
        r^2 -2r + 1 \\
        1 - r
        \end{bmatrix}
        \\[1em]
        \nonumber
        &= \frac{(W_1 + W_T)(1 - r) + (1 - r)^2 \sum\limits_{s=2}^{T-1}W_s}{(1-r)(1+r)}
        \\
        \nonumber
        &= \frac{\sum\limits_{t=1}^{T}W_t - r \sum\limits_{s=2}^{T-1} W_s}{1+r} \, .
    \end{align}
    Therefore,
    \begin{align}
        \nonumber
        W^\prime B^{-1} 1_T 1^\prime_T B^{-1} W &= (W^\prime B^{-1} 1_T)^2
        \\
        \nonumber
        &= \frac{ \left( \sum\limits_{t=1}^{T}W_t - r \sum\limits_{s=2}^{T-1} W_s \right)^2}{(1+r)^2}
        \\
        \label{eq:PDF_b_3cor}
        &= \frac{r^2 (\sum\limits_{s=2}^{T-1}W_s)^2 -2r \sum\limits_{s=2}^{T-1}W_s \sum\limits_{t=1}^{T}W_t + (\sum\limits_{t=1}^{T}W_t)^2}{(1+r)^2}
    \end{align}
    Finally, the form \eqref{eq:PDF_b_simpl} is obtained 
    by substituting  into \eqref{eq:PDF_b_quasi_simpl}  the expressions
    \eqref{eq:PDF_b_1cor}, \eqref{eq:PDF_b_2cor}, and \eqref{eq:PDF_b_3cor}.    
\end{proof}
It is important to note that equation \eqref{eq:PDF_b_simpl} is inexpensive to
evaluate, as it is just a combination of ratios and products of polynomials in
$b$. This allows us to adopt the following two-step strategy. We first perform a
grid search for the maximum of $\pi(\bsZ|b)$, as a function of $b$. To this end,
we compute \eqref{eq:PDF_b_simpl} over a dense sequence of values of $b$, say
$-1.99, -1.98, \dots, -0.01$, and set
\begin{equation*}
    b_0 = \arg \max_{b = -1.99, -1.98, \dots, -0.01} \pi(\bsZ \vert b) \, .
\end{equation*}
It is reasonable that the value of $b_0$ is close to the global maximum of
$\pi(\bsZ \vert b)$. Then, we use the L-BFGS-B algorithm to find the global
maximum starting from $b_0$. Although this optimization must be repeated at each
iteration of the Gibbs sampler, our method is fast due to the simplicity of
expression \eqref{eq:PDF_b_simpl}. In fact, as reported in
Section~\ref{sec:simStudy}, the Gibbs sampler is much faster than the MCMC-EM
optimizer. Furthermore, as reported in the Supplementary Material, our proposed
Gibbs sampler is also faster than the state-of-the-art represented by the Stan
implementation.


\section{Numerical Experiments}
\label{sec:exp}

The purpose of running the numerical experiments in this section is two-fold.
First, we examine the difference between the frequentist and Bayesian
inferences, which can be informative when the data has a modest size and we want
to gauge the influence of the prior. Second, we provide proof-of-concept for the
algorithms proposed and examine their performance on data generated under
different scenarios, including one in which the sampling error distribution is
misspecified. We also consider a comparison between the Bayesian inference
produced using the sampler we design here and the one using a generic Stan
implementation. The latter can be considered at this point to be state of the
art since, to our knowledge, no other MCMC sampler has been designed for this
problem, except the generic implementation of \citet{lele2007data}. 

In this paper's numerical experiments designed to compare algorithms, we
replaced the now deprecated WinBUGS with Stan, which is widely regarded as the
current state of the art off-the-shelf algorithm which, generally, is known to
exhibit superior efficiency compared to WinBugs.

\subsection{Simulations}
\label{sec:simStudy}

\subsubsection{Simulation scenarios}
The simulation study evaluates the performance of the proposed algorithms when
the sampling model is correct and when it is misspecified. Specifically, we
generated time series of varying lengths under the Poisson noise model and under
a model with sampling errors that have a negative binomial distribution. In
addition, we varied the model parameters to examine the algorithms performance
under different scenarios. One of the scenarios has parameter values close to
the estimates obtained in the Redstart data analysis. 

The data are fitted using the frequentist approach described in
Section~\ref{sec:freq}, and using the Bayesian methods described in
Section~\ref{sec:Bay}. In addition, Bayesian inference is produced using an
off-the-shelf Stan implementation.

We analyze data generated under eight simulation scenarios.  Scenario {\bf S1}
yields a moderate level of serial correlation given by $b^\dagger=-0.5$, while
setting {\bf S2} exhibits high levels of correlation given by $b^\dagger=-0.22$.
Based on the parameters estimates obtained from the Redstart data analysis, we
fixed $\theta^\dagger_1 = 2$ and $\theta^\dagger_2 = 0.22$, for both {\bf S1}
and {\bf S2}.   We simulated these settings considering time series of lengths
$T = 30$. The scenarios {\bf S3}  and {\bf S4}  use the same parameters as,
respectively, {\bf S1}  and {\bf S2}, but have $T = 100$. 

Scenarios {\bf S5} - {\bf S8} are produced using the same parameter values as
{\bf S1} - {\bf S4}, respectively, but with a misspecified sampling error model.
Specifically, the sampling error was simulated using a negative binomial
distribution with mean $\exp(Z_t)$ and variance $2\exp(Z_t)$. These values were
chosen to provide a realistic comparison with the Poisson model. Note that a
negative binomial with the mean and variance specified above corresponds to one
with a success probability of $0.5$ and the dispersion (size) parameter selected
so that the mean equals $ \exp(Z_t)$. As the probability of success $p$
approaches zero, the negative binomial distribution converges to a Poisson
distribution with the same mean. Thus, setting $p=0.5$ ensures that the model is
sufficiently different to illustrate the impacts of misspecification. For each
setting, we then computed the mean square error (MSE) and the coverage of the
95\% confidence and credible intervals, for every parameter in each scenario. 

The simulation analysis is conducted with 500 independent replicates, and for
the Bayesian methods, each replicate is based on an MCMC sample of size
$10,000$. In addition, we report the effective sample sizes obtained by our
sampler and Stan. Finally, we compute and report summaries of computation times.

\begin{figure}[ht]
\centering
    \begin{subfigure}[b]{0.5\textwidth}            
            \includegraphics[width=\textwidth]{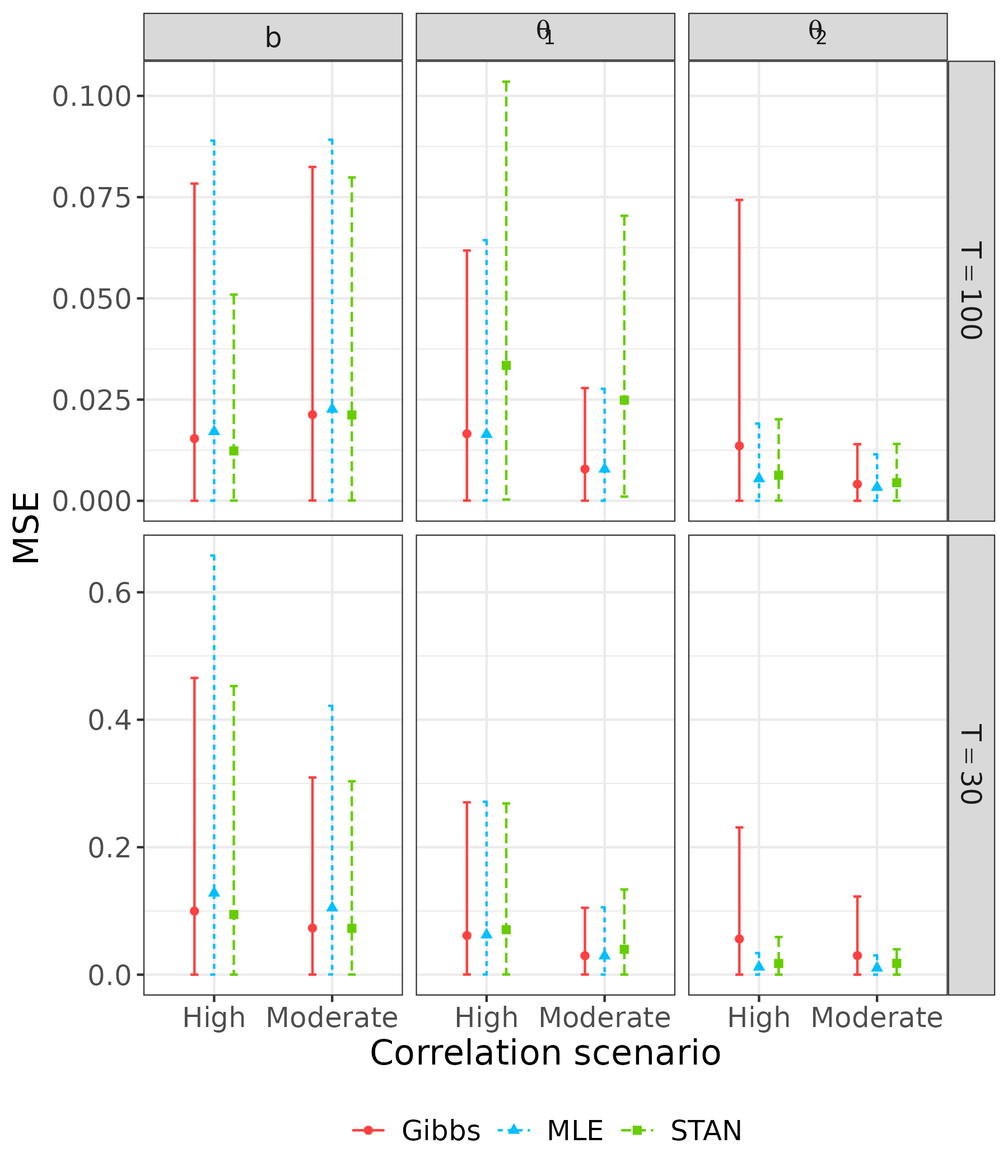}
            \caption{MSE}
            \label{fig:MSE1}
    \end{subfigure}%
    \begin{subfigure}[b]{0.5\textwidth}
            \centering
            \includegraphics[width=\textwidth]{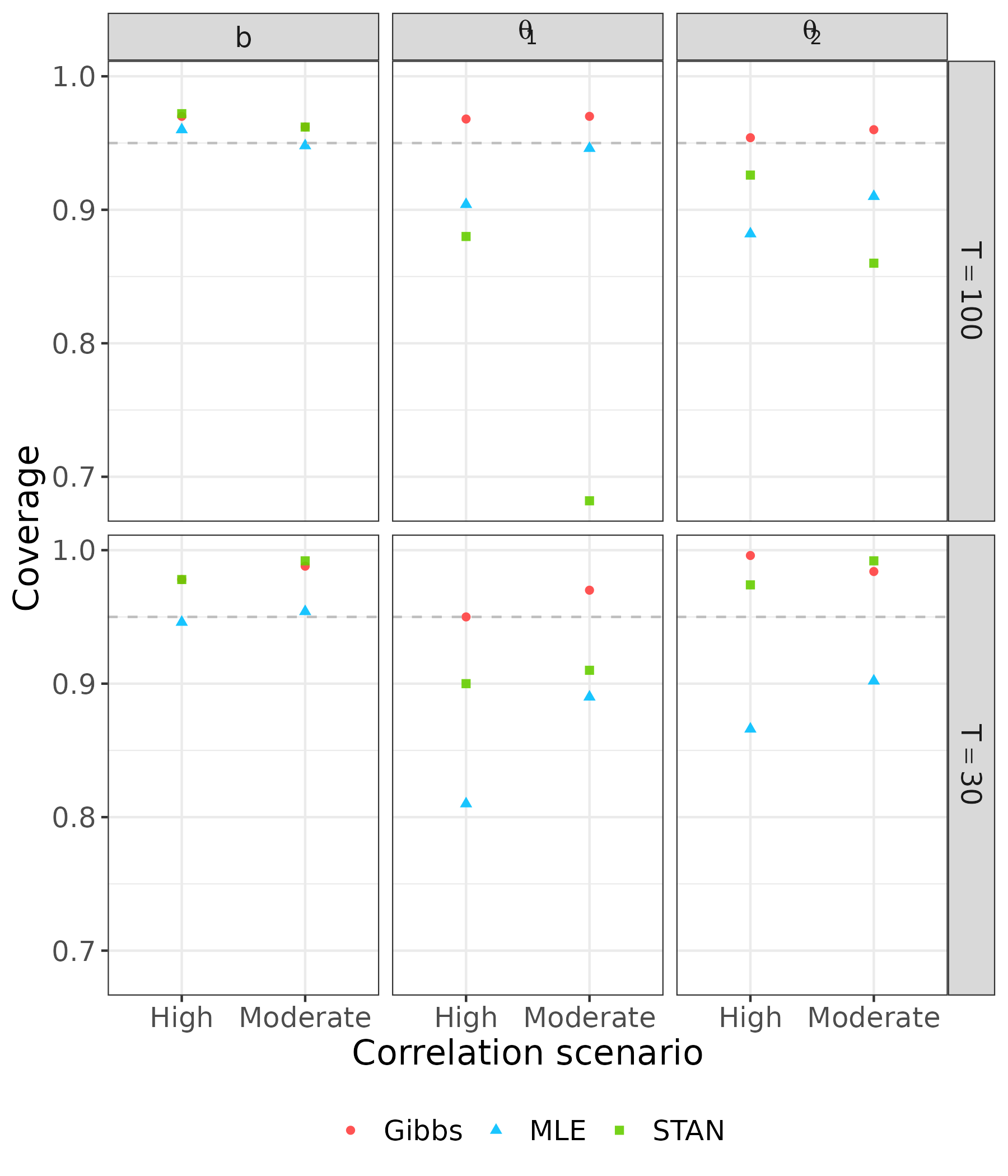}
            \caption{Coverage}
            \label{fig:cov1}
    \end{subfigure}
\caption{\textbf{Results for the correctly specified model with Poisson
distributed errors}: Mean square error and $95\%$ coverage by correlation
setting for each parameter and time series length $T$ for scenarios {\bf S1} -
{\bf S4}. Within each sub-figure, columns represent parameters and rows
represent the different time series length $T$. Results form the Gibbs sampler
are shown with points and solid lines results, results for the MLE approach are
indicated by triangles and fine dashed lines, and results for Stan are displayed
with squares and bold dashed lines. In the MSE plots (left panels), the central
points/triangles indicate the mean values, and the error bars represent the 5th
and 95th percentiles. The coverage plots (right panels) show the coverage
obtained using credible intervals approximated from Gibbs samples (red points),
Stan samples (green squares), and using Wald confidence intervals obtained using
the MLE and is asymptotic variance produced by MCMC-EM.}
\label{Fig:correctModel}
\end{figure}

\begin{figure}[ht]
\centering
 \begin{subfigure}[b]{0.5\textwidth}            
            \includegraphics[width=\textwidth]{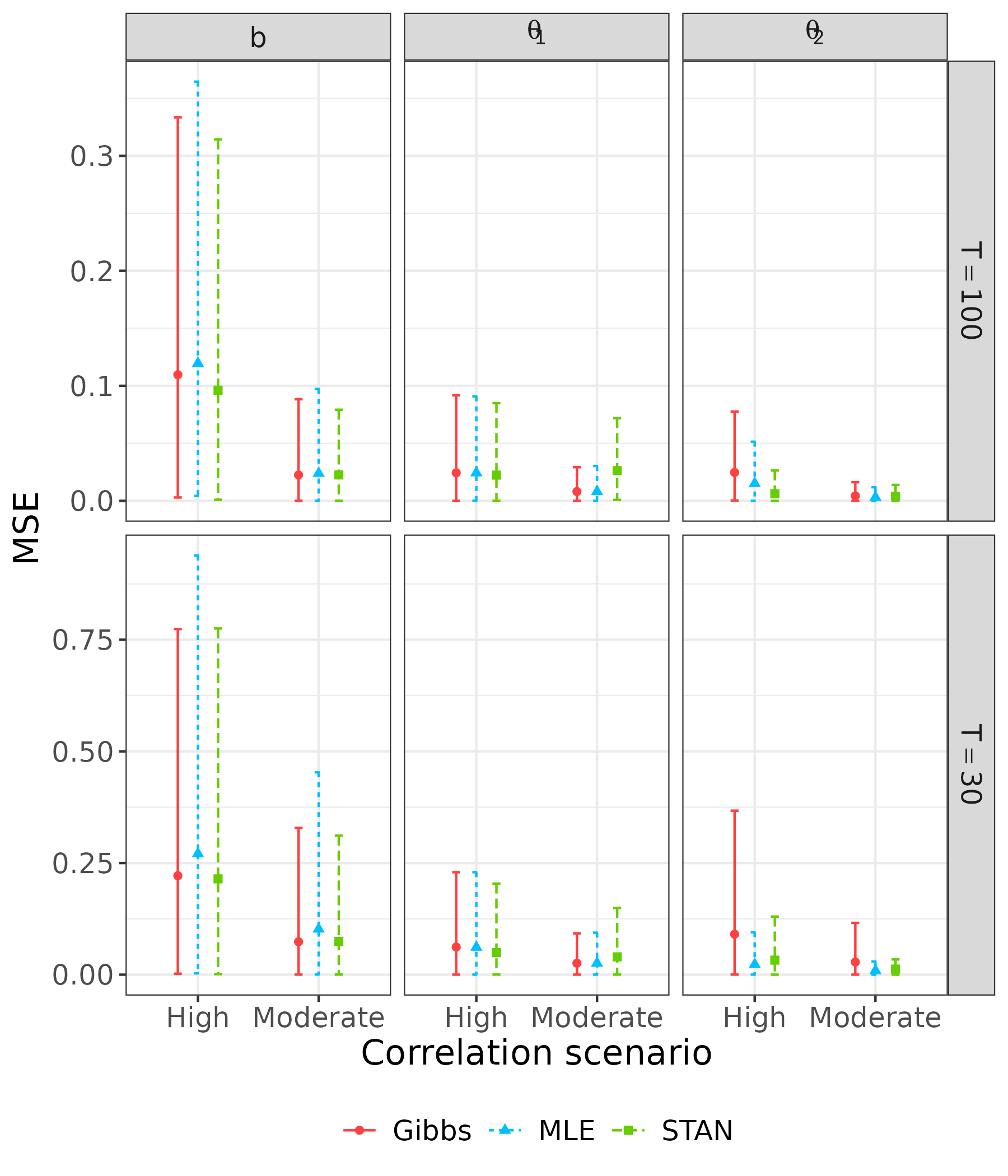}
            \caption{MSE}
            \label{fig:MSE2}
    \end{subfigure}%
  \hfill
    \begin{subfigure}[b]{0.5\textwidth}
            \centering
            \includegraphics[width=\textwidth]{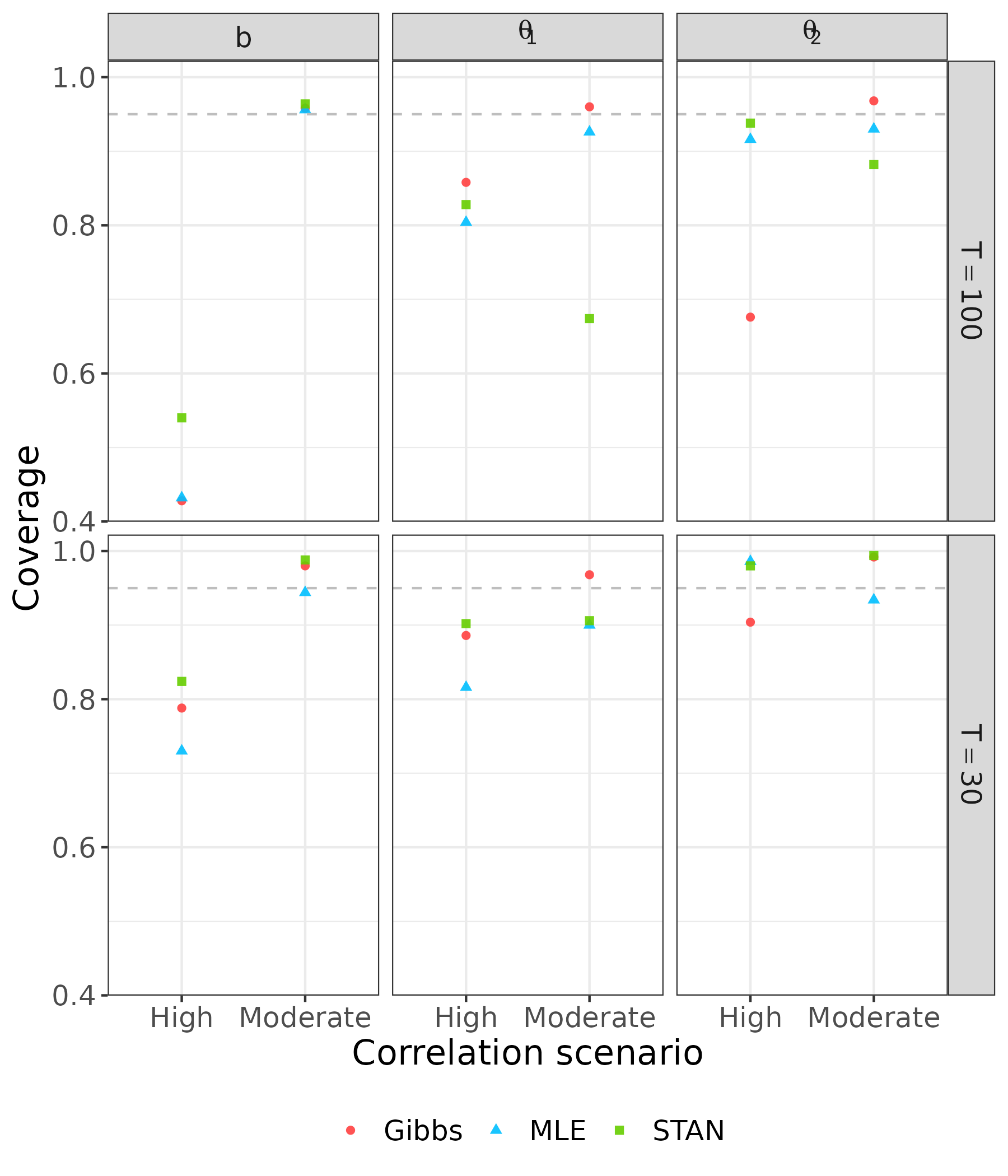}
            \caption{Coverage}
            \label{fig:cov2}
    \end{subfigure}
\caption{\textbf{Results for the misspecified model with Negative binomial
distributed errors}: Mean square error and $95\%$ coverage by correlation
scenario for each parameter and time series length $T$ for scenarios {\bf S5} -
{\bf S8}. Within each sub-figure, columns represent parameters and rows
represent the different time series length $T$. Results form the Gibbs sampler
are shown with points and solid lines results, results for the MLE approach are
indicated by triangles and fine dashed lines, and results for Stan are displayed
with squares and bold dashed lines. In the MSE plots (left panels), the central
points/triangles indicate the mean values, and the error bars represent the 5th
and 95th percentiles. The coverage plots (right panels) show the coverage
obtained using credible intervals approximated from Gibbs samples (red points),
Stan samples (green squares), and using Wald confidence intervals obtained using
the MLE and is asymptotic variance produced by MCMC-EM.}
 \label{Fig:misspModel}

\end{figure}

\subsubsection{Simulation results}

As expected, the performance of all three algorithms improves as the time series
length increases. Figure \ref{Fig:correctModel} presents the results for the
correctly specified model with Poisson distributed errors. In most cases, the
MSE values across the three computational methods are similar. The uncertainty
is generally higher for the estimates of the parameter $b$, lower for
$\theta_1$, and even lower for $\theta_2$. An exception occurs with Stan when
$T=100$, where the uncertainty for $\theta_1$ is greater than for $b$.
Furthermore, for $\theta_2$, the Gibbs sampler produced estimates with greater
uncertainty compared to the other two algorithms. In terms of coverage, all
three inference methods yield values close to $0.95$ for the parameter $b$.
However, the MLE approach shows lower coverage, particularly when $T=30$ and the
correlation is high. This is likely due to the sample size being too small for
the asymptotic variance estimate to be reliable.   When $T=100$, the Gibbs
sampler demonstrates the most accurate coverage across all parameters and
correlation scenarios.  

The results for the misspecified case are shown in Figure \ref{Fig:misspModel}.
In this setting, the performance of the algorithms differs noticeably between
the high and moderate correlation scenarios. When the correlation is lower, the
three algorithms exhibit improved performance, with lower MSE, reduced
uncertainty, and more accurate coverage across all parameters and both sample
sizes. Again, a higher uncertainty is observed for the estimates of $b$ and
$\theta_1$, and a lower uncertainty is observed for $\theta_2$. Moreover, when
$T=100$, the Gibbs sampler provides the most accurate coverage for all
parameters. 

\begin{table}[ht]
    \centering
    \begin{tabular}{ccccc}\toprule
       &  1st Quantile&  Mean& Median& 3rd Quantile\\
        \midrule
        \bf{Gibbs}&  1.33&  1.33&  1.33& 1.34\\
        \bf{MLE} &  9.21&  11.82&  11.23& 13.38\\
        \bf{Stan}&  6.98&  7.66&  7.09 & 7.31\\
        \bottomrule
    \end{tabular}
    \caption{Summary statistics of the computational times of each algorithm (in minutes) under scenario {\bf S4}}
    \label{tab:computational_times}
\end{table}

The computational times for scenario {\bf S4} are reported in Table
\ref{tab:computational_times}. The MLE approach was the slowest, taking between
two and nine times longer than the other two. The Gibbs sampler outperforms Stan
in terms of speed, with a mean computation time that is approximately five times
shorter. The other scenarios exhibited similar patterns and are presented in the
Supplementary Material. Finally, a comparison of the effective sample sizes for
the three parameters across the two Bayesian inference methods is also provided
there. In all cases, the Gibbs sampler outperformed Stan, yielding effective
sample sizes per second that were around 1.5 to 3.5 times larger. This is not
surprising, since the algorithm presented here has been designed specifically
for this problem, while Stan is a sampling algorithm which is able to tackle a
wide variety of models. Due to its general-purpose nature, Stan is not tailored
to exploit the specific characteristic of a given target. Finally, it is
important to note that if the algorithms are run for a much larger number of
iterations, both the Gibbs sampler and Stan will yield similar conclusions about
the posterior. However, the number of iterations required for the two algorithms
to produce similar results may be computationally prohibitive. In fact, the only
theoretical guarantee is that they asymptotically converge to the same result as
the number of iterations approaches infinity. Since the number of iterations is
finite, the differences reported arise from the different mixing times of the
two algorithms, which are particularly relevant for shorter runs.

\subsection{Real data analysis} \label{realData}

We analyze here the data set of American Redstart counts that was previously
discussed by \cite{Lele} and \cite{Dennis_2006}. This data set is recorded with
the number 0214332808636 in the North American Breeding Bird Survey
\citep{peterjohn1994north,robbins1986breeding} and contains the number of
specimens observed from 1966 to 1995 at a survey location. We fit the Gompertz
model with Poisson-distributed errors to this dataset using both the Gibbs
sampler and the MLE approach proposed here. Stan was also used as a baseline for
comparison within the Bayesian framework.

\begin{figure}
    \centering
    \includegraphics[width=0.75\linewidth]{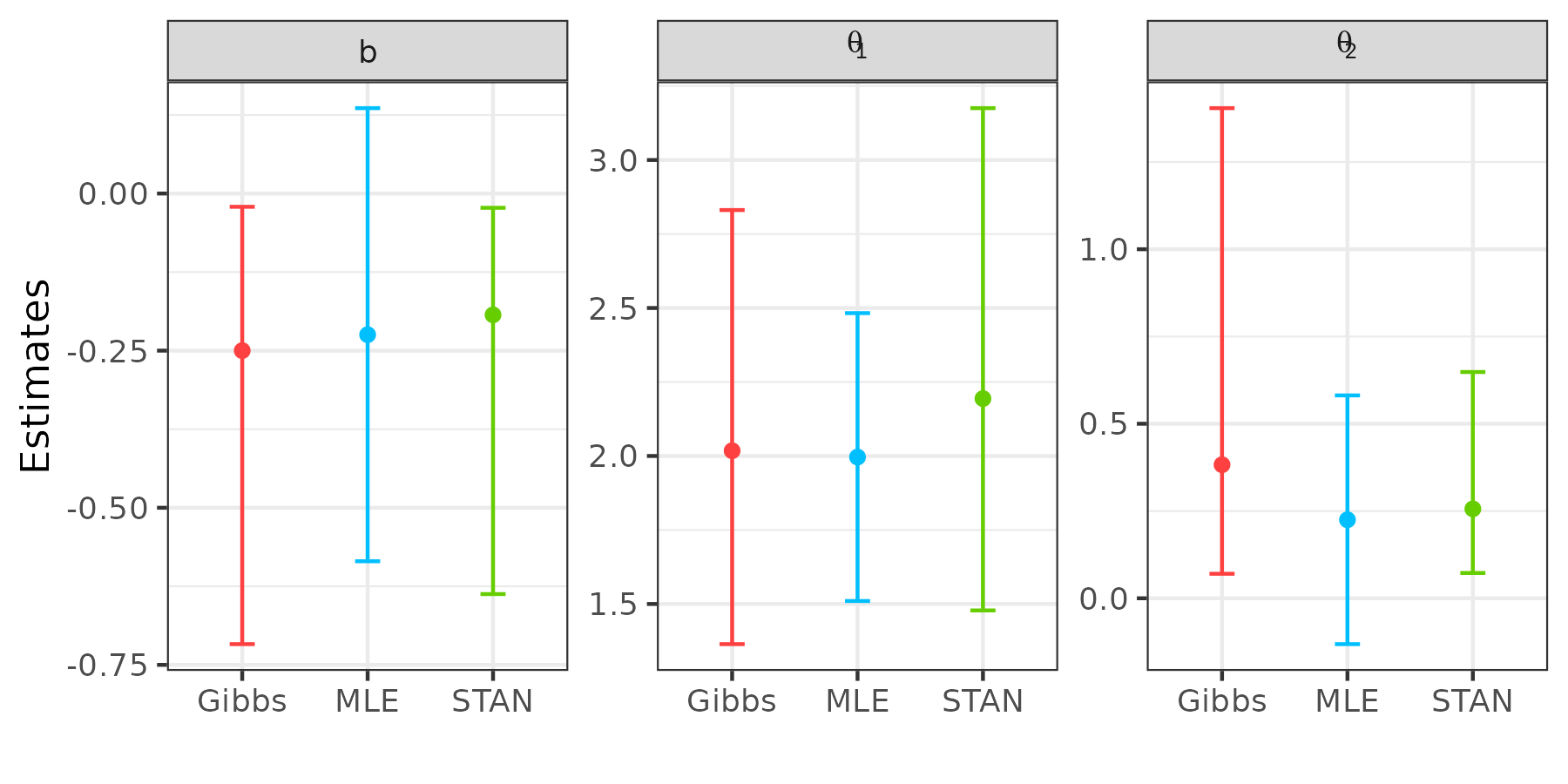}
    \caption{Point estimates for the three parameters obtained using each inference method. For the two Bayesian methods, the error bars represent the 95\% credible intervals, while for the MLE, they indicate the Wald 95\%  confidence intervals .}
    \label{fig:estimates_RDA}
\end{figure}

\begin{figure}
    \centering
    \includegraphics[width=0.75\linewidth]{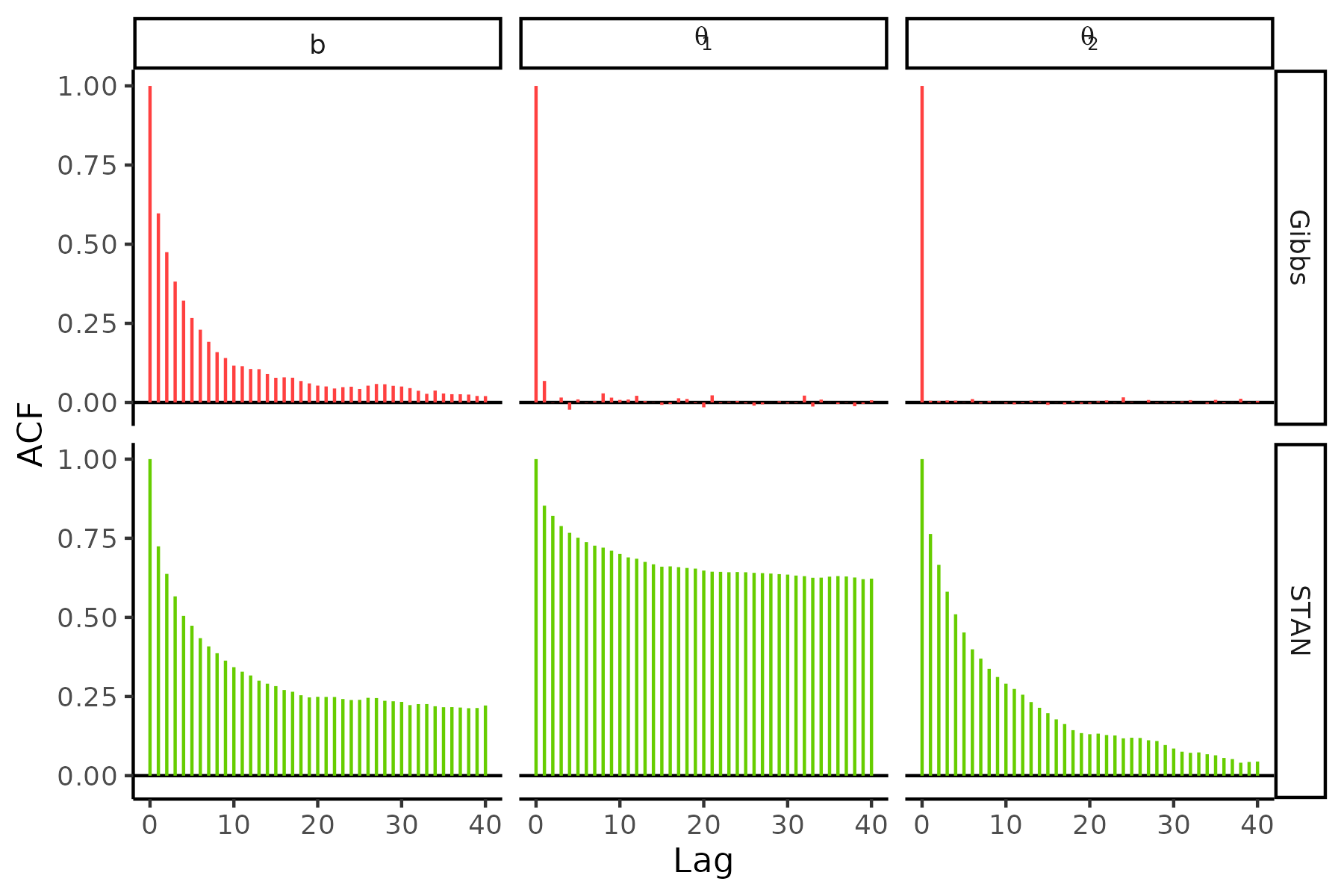}
    \caption{Autocorrelation functions of the posterior samples for the three parameters under both Bayesian inference methods. }
    \label{fig:acf_RDA}
\end{figure}

We computed point estimates for the three parameters, along with the $95\%$
credible intervals for the Bayesian methods and the $95\%$ confidence intervals
based on Gaussian approximation for the MLE approach (Figure
\ref{fig:estimates_RDA}). For the two Bayesian methods, we also calculated the
effective sample size and autocorrelation function (Figure \ref{fig:acf_RDA})
for each parameter (Table \ref{tab:eff_rd}).  

Although the Gibbs sampler shows greater uncertainty in the estimation of
$\theta_2$, the point estimates across the three methods remain comparable.
However, when examining the effective sample size and autocorrelation functions
of the posterior distributions for the two Bayesian methods, the Gibbs sampler
demonstrates superior performance. The effective sample size for $\theta_2$ is
nearly $20$ times as large with the Gibbs sampler, six times larger for $b$, and
increases for $\theta_1$, from only $30$ in Stan to more than $8200$ with the
Gibbs sampler.

\begin{table}
    \centering
    \begin{tabular}{cccc}
            \toprule
         &  b&  $\theta_1$&  $\theta_2$\\
         \midrule
       \textbf{Gibbs} &  1106.8&  8239.5& 10000\\
        \textbf{Stan} &  209.4&  30.3& 583.4\\
        \bottomrule
    \end{tabular}
    \caption{Effective sample sizes for the three parameters from the analysis of the American Redstart for both Bayesian inference methods}
    \label{tab:eff_rd}
\end{table} 

\section{Discussion}\label{sec:disc}

We develop full-likelihood-based inference within the frequentist and Bayesian
paradigms for the Gompertz model with Poisson sampling errors. The proposed
approaches remove the need to consider pseudo-likelihood methods that mitigate
computational challenges at the price of reducing the information provided by
the data. 

In our future work, we would like to investigate whether similar developments
can be produced to modified versions of the model considered here. The latter
can be created by modifying the growth curve by adding parameters that allow
curvature and long-term behavior \citep{asadi2023generalized} or when the
population dynamics is determined by a stochastic differential equation as in
\cite{donnet2010bayesian}.

\section*{Acknowledgements}

The first author was supported by MUR - Prin 2022 - Grant no. 2022FJ3SLA, funded
by the European Union – Next Generation EU. The third author was supported by
NSERC of Canada discovery grant RGPIN-2024-04506. The authors thank Dr.\ Monica
Alexander and Dr.\ Vianey Leos-Barajas  for comments that have improved the
paper.

\section*{Code}
The developed methods are available in a new R package: \texttt{gse}. The new R
package and the scripts of Section~\ref{sec:exp} are available online on GitHub
at the following link: \texttt{https://github.com/sofiar/GMLossF}

\bibliography{references.bib}

@article{baione2018pricing,
	author = {Baione, Fabio and Levantesi, Susanna},
	journal = {North American Actuarial Journal},
	number = {2},
	pages = {270--288},
	publisher = {Taylor \& Francis},
	title = {Pricing critical illness insurance from prevalence rates: Gompertz versus {We}ibull},
	volume = {22},
	year = {2018}}

@article{tai2018models,
	author = {Tai, Tzu Han and Noymer, Andrew},
	journal = {Population Ecology},
	number = {1-2},
	pages = {171--184},
	publisher = {Wiley Online Library},
	title = {Models for estimating empirical {Gompertz} mortality: With an application to evolution of the {Gompertzian} slope},
	volume = {60},
	year = {2018}}

@article{donnet2010bayesian,
  title={Bayesian analysis of growth curves using mixed models defined by stochastic differential equations},
  author={Donnet, Sophie and Foulley, Jean-Louis and Samson, Adeline},
  journal={Biometrics},
  volume={66},
  number={3},
  pages={733--741},
  year={2010},
  publisher={Oxford University Press}
}

@article{asadi2023generalized,
  title={A generalized {G}ompertz growth model with applications and related birth-death processes},
  author={Asadi, Majid and Di Crescenzo, Antonio and Sajadi, Farkhondeh A and Spina, Serena},
  journal={Ricerche di Matematica},
  volume={72},
  number={2},
  pages={1--36},
  year={2023},
  publisher={Springer}
}

@article{alexander2017flexible,
	author = {Alexander, Monica and Zagheni, Emilio and Barbieri, Magali},
	journal = {Demography},
	number = {6},
	pages = {2025--2041},
	publisher = {Duke University Press},
	title = {A flexible {B}ayesian model for estimating subnational mortality},
	volume = {54},
	year = {2017}}

@article{butt2004application,
	author = {Butt, Zoltan and Haberman, Steven},
	journal = {ASTIN Bulletin: The Journal of the IAA},
	number = {1},
	pages = {175--197},
	publisher = {Cambridge University Press},
	title = {Application of frailty-based mortality models using generalized linear models},
	volume = {34},
	year = {2004}}

@article{staples2004estimating,
	author = {Staples, David F and Taper, Mark L and Dennis, Brian},
	journal = {Ecology},
	number = {4},
	pages = {923--929},
	publisher = {Wiley Online Library},
	title = {Estimating population trend and process variation for PVA in the presence of sampling error},
	volume = {85},
	year = {2004}}

@article{mcculloch1994maximum,
  title={Maximum likelihood variance components estimation for binary data},
  author={McCulloch, Charles E},
  journal={Journal of the American Statistical Association},
  volume={89},
  number={425},
  pages={330--335},
  year={1994},
  publisher={Taylor \& Francis}
}

@article{chan1995monte,
  title={Monte {Carlo EM} estimation for time series models involving counts},
  author={Chan, Kung-Sik and Ledolter, Johannes},
  journal={Journal of the American Statistical Association},
  volume={90},
  number={429},
  pages={242--252},
  year={1995},
  publisher={Taylor \& Francis}
}

@article{booth1999maximizing,
  title={Maximizing generalized linear mixed model likelihoods with an automated {Monte Carlo EM} algorithm},
  author={Booth, James G and Hobert, James P},
  journal={Journal of the Royal Statistical Society Series B: Statistical Methodology},
  volume={61},
  number={1},
  pages={265--285},
  year={1999},
  publisher={Oxford University Press}
}

@article{wei1990monte,
  title={A {Monte Carlo} implementation of the {EM} algorithm and the poor man's data augmentation algorithms},
  author={Wei, Greg CG and Tanner, Martin A},
  journal={Journal of the American statistical Association},
  volume={85},
  number={411},
  pages={699--704},
  year={1990},
  publisher={Taylor \& Francis}
}

@article{linden2009estimating,
	author = {Lind{\'e}n, Andreas and Knape, Jonas},
	journal = {Oikos},
	number = {5},
	pages = {675--680},
	publisher = {Wiley Online Library},
	title = {Estimating environmental effects on population dynamics: consequences of observation error},
	volume = {118},
	year = {2009}}

@article{Lele,
	author = {Subhash R. Lele},
	journal = {Ecology},
	number = {1},
	pages = {189--202},
	publisher = {Ecological Society of America},
	title = {Sampling Variability and Estimates of Density Dependence: A Composite-Likelihood Approach},
	urldate = {2022-09-20},
	volume = {87},
	year = {2006}}

@article{gompertz1825xxiv,
  author={Gompertz, Benjamin},
title={On the nature of the function expressive of the law of human mortality, and on a new mode of determining the value of life contingencies. {In} a letter to {Francis Baily, Esq. FRS} },
  journal={Philosophical transactions of the Royal Society of London},
  number={115},
  pages={513--583},
  year={1825},
  publisher={The Royal Society London}
}

@book{newman_book,
	author = {Newman, Ken and Buckland, Steve and Morgan, Byron and King, Ruth and Borchers, David and Cole, Diana and Besbeas, Panagiotis and Gimenez, Olivier and Thomas, Len},
	title = {Modelling Population Dynamics: Model Formulation, Fitting and Assessment using State-Space Methods},
    publisher= {Springer},
	year = {2014}}

@article{Dennis_2006,
	author = {Dennis, Brian and Ponciano, Jos{\'e} Miguel and Lele, Subhash R. and Taper, Mark L. and Staples, David F.},
	journal = {Ecological Monographs},
	number = {3},
	pages = {323-341},
	title = {ESTIMATING DENSITY DEPENDENCE, PROCESS NOISE, AND OBSERVATION ERROR},
	volume = {76},
	year = {2006}}

@article{Methe_2021,
	author = {Auger-M{\'e}th{\'e}, Marie and Newman, Ken and Cole, Diana and Empacher, Fanny and Gryba, Rowenna and King, Aaron A. and Leos-Barajas, Vianey and Mills Flemming, Joanna and Nielsen, Anders and Petris, Giovanni and Thomas, Len},
	eprint = {https://esajournals.onlinelibrary.wiley.com/doi/pdf/10.1002/ecm.1470},
	journal = {Ecological Monographs},
	number = {4},
	pages = {e01470},
	title = {A guide to state--space modeling of ecological time series},
	volume = {91},
	year = {2021}}

@article{Hostetler_2015,
	author = {Hostetler, Jeffrey A. and Chandler, Richard B.},
	eprint = {https://esajournals.onlinelibrary.wiley.com/doi/pdf/10.1890/14-1487.1},
	journal = {Ecology},
	number = {6},
	pages = {1713-1723},
	title = {Improved state-space models for inference about spatial and temporal variation in abundance from count data},
	volume = {96},
	year = {2015}}

@article{tjovre_2017,
	author = {Tj{\o}rve, Kathleen M. C. AND Tj{\o}rve, Even},
	journal = {PLOS ONE},
	month = {06},
	number = {6},
	pages = {1-17},
	publisher = {Public Library of Science},
	title = {The use of {Gompertz} models in growth analyses, and new {Gompertz}-model approach: An addition to the Unified-{Richards} family},
	volume = {12},
	year = {2017}}

@article{Winsor1932-ii,
	address = {United States},
	author = {Winsor, C P},
	journal = {Proc Natl Acad Sci U S A},
	language = {en},
	month = jan,
	number = 1,
	pages = {1--8},
	title = {The {Gompertz} Curve as a Growth Curve},
	volume = 18,
	year = 1932}

@article{dempster1977,
  title={Maximum likelihood from incomplete data via the {EM} algorithm},
  author={Dempster, A.P. and Laird, N.M. and Rubin, D.B.},
  journal={Journal of the royal statistical society: series B (methodological)},
  volume={39},
  number={1},
  pages={1--22},
  year={1977},
  publisher={Wiley Online Library}
}

@article{caffo2005,
  title={Ascent-based {Monte Carlo} expectation--maximization},
  author={Caffo, B.S. and Jank, W. and Jones, G.L.},
  journal={Journal of the Royal Statistical Society Series B: Statistical Methodology},
  volume={67},
  number={2},
  pages={235--251},
  year={2005},
  publisher={Oxford University Press}
}

@article{louis1982,
  title={Finding the observed information matrix when using the {EM} algorithm},
  author={Louis, T.A.},
  journal={Journal of the Royal Statistical Society Series B: Statistical Methodology},
  volume={44},
  number={2},
  pages={226--233},
  year={1982},
  publisher={Oxford University Press}
}

@article{peterjohn1994north,
  title={The {North American} breeding bird survey},
  author={Peterjohn, B},
  journal={Birding},
  volume={26},
  number={6},
  pages={386--398},
  year={1994}
}

@techreport{robbins1986breeding,
  title={The {Breeding Bird Survey}: its first fifteen years, 1965-1979},
  author={Robbins, Chandler S and Bystrak, Danny and Geissler, Paul H},
  year={1986},
institution = {US Fish and Wildlife Service}
}

@book{hamilton1994time,
  title     = {Time Series Analysis},
  author    = {Hamilton, J.D.},
  year      = {1994},
  publisher = {Princeton University Press},
  address   = {Princeton, NJ}
}

@article{lele2007data,
  title={Data cloning: easy maximum likelihood estimation for complex ecological models using {B}ayesian {M}arkov chain {M}onte {C}arlo methods},
  author={Lele, Subhash R and Dennis, Brian and Lutscher, Frithjof},
  journal={Ecology letters},
  volume={10},
  number={7},
  pages={551--563},
  year={2007},
  publisher={Wiley Online Library}
}

\end{document}


\section*{Supplement: Extra Figures and Tables}
\begin{figure}[H]
    \centering
 \begin{subfigure}[b]{0.5\textwidth}            
            \includegraphics[width=\textwidth]{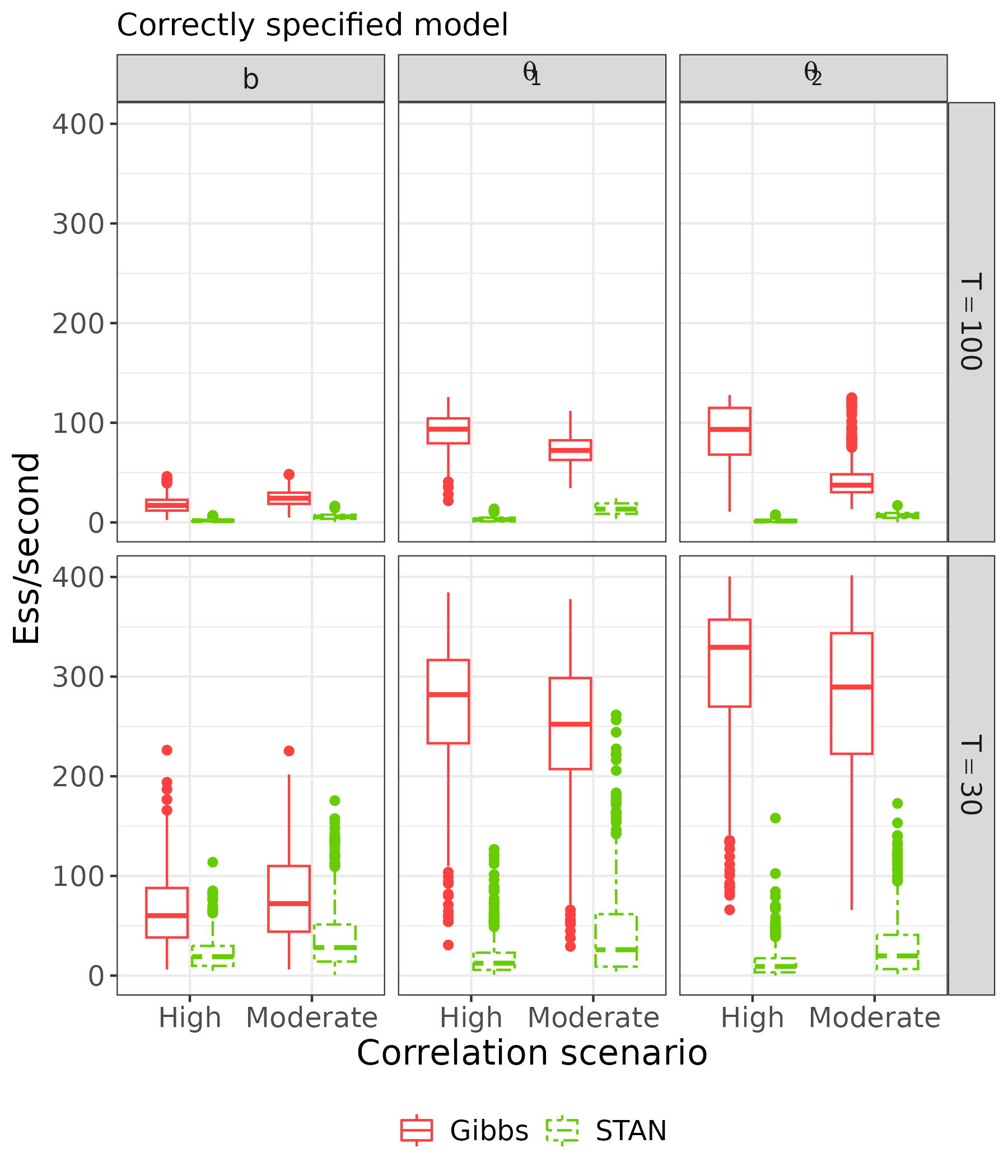}
            \label{fig:MSE1}
    \end{subfigure}%
    \begin{subfigure}[b]{0.5\textwidth}
            \centering
            \includegraphics[width=\textwidth]{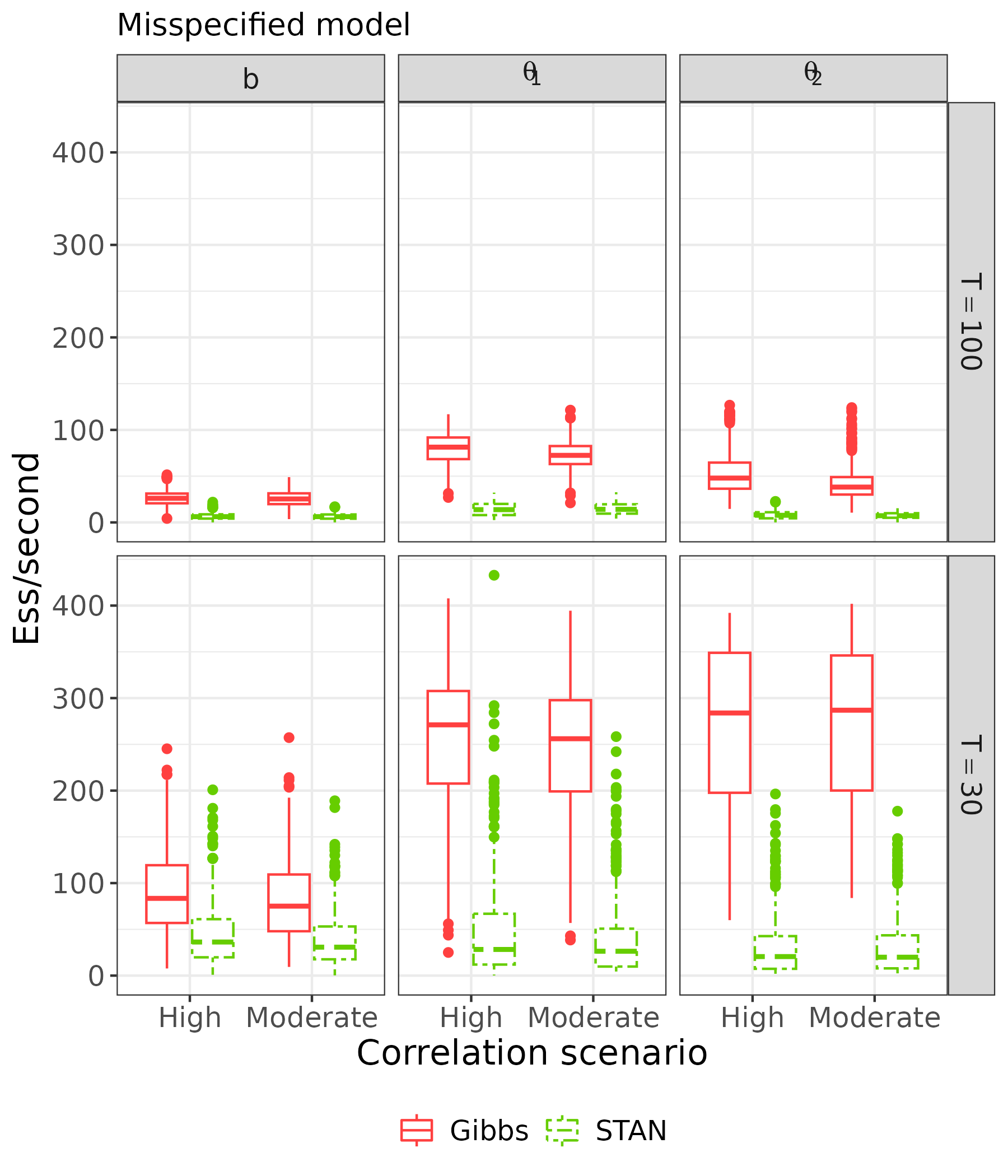}
            \label{fig:cov1}
    \end{subfigure}
    \caption{Effective sample size per second obtained from the simulation analysis for both Bayesian inference methods for each parameter and time series length $T$.  Results form the Gibbs sampler are shown with solid red lines, while those from Stan are displayed with green dashed lines.}
    \label{fig:enter-label}
\end{figure}

\begin{table}[h]
    \centering
    \begin{tabular}{cccccc}\toprule
         Correctly specified model &&&&&\\
        \midrule
          &  1st Quantile&  Mean& Median& 3rd Quantile& Scenario\\
        \bf{Gibbs}& 0.44& 0.45 &0.45 &0.46 & {\bf S1} \\
        \bf{MLE} & 3.43 &4.82  & 4.11 &5.41&{\bf S1}\\
        \bf{Stan}& 0.57 &0.97  &0.76  &0.97& {\bf S1}\\
        \midrule
        \bf{Gibbs}&0.43 & 0.44 & 0.44&0.44& {\bf S2} \\
        \bf{MLE} &3.04  & 4.16 & 3.48 &4.55&{\bf S2}\\
        \bf{Stan}& 0.53 & 0.7 & 0.57 &0.8& {\bf S2}\\
        \midrule
        \bf{Gibbs}& 1.32& 1.34 & 1.33&1.37& {\bf S3} \\
        \bf{MLE} & 11.46 & 14.5 & 13.58 &16.87&{\bf S3}\\
        \bf{Stan}& 8.77 & 12.7 & 11.82 &14.87& {\bf S3}\\
        \bottomrule
    \end{tabular}
    \caption{ Summary statistics of the computational times (in minutes) under scenarios {\bf S1-S3}}
    \label{tab:computational_times1}
\end{table}

\begin{table}[h]
    \centering
    \begin{tabular}{cccccc}\toprule
         Misspecified model &&&&&\\
        \midrule
          &  1st Quantile&  Mean& Median& 3rd Quantile& Scenario\\
        \bf{Gibbs}& 0.44& 0.45 &0.45 & 0.46& {\bf S1} \\
        \bf{MLE} & 3.02 &3.9 & 3.49 & 4.05 &{\bf S1}\\
        \bf{Stan}& 0.53 & 0.67 & 0.58& 0.72& {\bf S1}\\
        \midrule
        \bf{Gibbs}& 0.43&  0.44& 0.44& 0.45& {\bf S2} \\
        \bf{MLE} & 3.04 & 3.9& 3.59 & 4.36&{\bf S2}\\
        \bf{Stan}& 0.53 & 0.69 &0.56 &0.76 &{\bf S2}\\
       \midrule
        \bf{Gibbs}& 1.33& 1.36 &1.37 &1.39 & {\bf S3} \\
        \bf{MLE} & 10.06 & 12.12& 11.59 &13.97 &{\bf S3}\\
        \bf{Stan}& 7.04 & 7.71 & 7.15& 3.37&{\bf S3}\\
         \midrule
        \bf{Gibbs}&1.33 & 1.34 &1.34 &1.35 & {\bf S4} \\
        \bf{MLE} & 9.49 & 11.83&  11.24& 13.29&{\bf S4}\\
        \bf{Stan}& 7 & 7.66 &7.11 & 7.36&{\bf S4}\\

      \bottomrule
    \end{tabular}
    \caption{ Summary statistics of the computational times (in minutes) under scenarios {\bf S1-S4}}
    \label{tab:computational_times2}
\end{table}